# The tetragonal-cubic transition of davemaoite: Implications for lower mantle seismic anomalies

Yoshiyuki Okuda[1,2,*], Bin Chen[1,*], Juliana Peckenpaugh[1], Keng-Hsien Chao[1], Saori Kawaguchi-Imada[3], Hirokazu Kadobayashi[3], Zhenxian Liu[4], Dongzhou Zhang[5]

[1] *Hawai'i Institute of Geophysics and Planetology, University of Hawai'i at Manoa, Honolulu, Hawai'i 96822, USA*

[2] *Department of Earth and Planetary Sciences, Institute of Science Tokyo, Meguro, Tokyo 152-8551, Japan*

[3] *SPring-8, Japan Synchrotron Radiation Research Institute, Sayo, Hyogo 679-5198, Japan*

[4] *National Synchrotron Light Source II, Brookhaven National Laboratory, Upton, New York 11973, USA*

[5] *Center for Advanced Radiation Sources, The University of Chicago, Chicago, Illinois 60637, USA*

*Corresponding author

*E-mail: yokuda@hawaii.edu (Y. Okuda), binchen@hawaii.edu (B. Chen)

## ABSTRACT

Davemaoite ($CaSiO_3$ perovskite) is the third most abundant mineral in Earth's lower mantle and a dominant phase in subducted oceanic crust. Its crystal structure is distorted (tetragonal or orthorhombic) at ambient conditions but is considered to transform to cubic at high temperatures. Previous experiments reported low, nearly pressure-independent transition temperatures (~600 K), demonstrating a long-standing discrepancy with theoretical predictions that mostly exceed 1000 K. Here, we determine the phase stability and thermal equation of state of $CaSiO_3$ davemaoite and its titanium-bearing solid solution [$Ca(Si_{0.75},Ti_{0.25})O_3$] under simultaneous high-pressure and high-temperature conditions using a laser-heated diamond anvil cell combined with synchrotron X-ray diffraction. We find that the tetragonal-to-cubic transition occurs at substantially higher temperatures than previously reported, with the titanium substitution further stabilizing the tetragonal phase and shifting the transition boundary to even higher temperatures. These findings indicate that $CaSiO_3$ davemaoite in subducted oceanic crust likely undergoes its ferroelastic transition in the mid-lower mantle, whereas Ti-rich davemaoite may remain tetragonal throughout most of the lower mantle, transforming to cubic near the core–mantle boundary. Our results demonstrate that the compositionally dependent phase behaviour of davemaoite can account for the seismic anomalies observed in both the mid-lower mantle and the lowermost mantle.

## Highlights

- Davemaoite transforms to cubic at much higher temperatures than previously reported
- Titanium stabilizes tetragonal davemaoite and raises its transition temperature
- Davemaoite's tetragonal-cubic transition may explain lower-mantle seismic heterogeneity

## 1. Introduction

Subducted oceanic lithosphere (slabs) delivers a considerable amount of mid-ocean ridge basalt (MORB) into the Earth's lower mantle. Portions of the subducted MORB crust may detach from the slab (Nakagawa and Tackley, 2004) and accumulate at the top of the lower mantle (Irifune and Ringwood, 1993), forming chemically distinct layers that have been linked to seismic anomalies associated with slow shear velocities (Gréaux et al., 2019) and elevated electrical conductivity anomalies (Inada et al., 2025). With continued descent, detached MORB or entire slabs may eventually sink through the lower mantle to the core-mantle boundary (CMB), causing chemical heterogeneity throughout the lower mantle (Feng et al., 2021).

A major constituent of subducted MORB is $CaSiO_3$ perovskite, officially named davemaoite in 2021 (Tschauner et al., 2021), which accounts for ~23 vol% of subducted oceanic crust (Hirose et al., 2005). Davemaoite is among the slowest seismic phases in the lower mantle and therefore plays a key role in making subducted MORB seismically slower than the surrounding mantle (Kudo et al., 2012; Zhou et al., 2025). It has been proposed as a major contributor to mid-mantle seismic scatterers observed beneath subduction zones, where shear-wave velocity reductions of a few to ~12% are reported (Feng et al., 2021; Kaneshima, 2016; Li and Yuen, 2014; Niu, 2014), and as a potential component of the large low shear velocity provinces (LLSVPs) above the CMB (Garnero et al., 2016; Thomson et al., 2019; Zhou et al., 2025).

At room temperature, pure $CaSiO_3$ davemaoite (Ca-davemaoite) crystallizes in a distorted tetragonal perovskite structure, but it transforms at high temperature into a cubic phase through a ferroelastic transition (Komabayashi et al., 2007). This transition is accompanied by spontaneous strain and elastic softening, which can substantially reduce seismic velocities in davemaoite and, consequently, in subducted MORB. Despite its significance for mantle seismic properties, experimental data on the tetragonal-cubic transition in davemaoite remain scarce. Previous measurements using an externally heated diamond anvil cell (EHDAC) (Komabayashi et al., 2007; Kurashina et al., 2004) suggested that the transition occurs near 600 K at lower mantle pressures. EHDAC studies identified the phase boundary by monitoring the (200) diffraction peak, which appears as a single peak in the cubic phase and splits into (200) and (002) peaks in the tetragonal structure. Their results imply that Ca-davemaoite is cubic throughout the lower mantle. In contrast, *ab initio* calculations predict much higher transition temperatures, on the order of 1200–2000 K at lower mantle pressures (Adams and Oganov, 2006; Stixrude et al., 2007, 1996; Thomson et al., 2019; Zhang et al., 2025), though some report values closer to experimental observations (Sagatova et al., 2021; Wu et al., 2024). Recent theoretical work has suggested that closely spaced tetragonal peaks can merge into an apparently single peak at high temperatures, which may underestimate the transition temperature (Zhang et al., 2025). Resolving this discrepancy requires *in situ* structural characterization at high pressures and temperatures (*P*–*T*) in a laser-heated diamond anvil cell (LHDAC), which can achieve >2000 K —temperatures beyond the reach of EHDAC techniques (Okuda et al., 2021; Wang et al., 2023). Yet no LHDAC studies have directly measured this transition boundary (Ono et al., 2004; Shim et al., 2000).

Impurities also play a critical role in controlling the crystal structure of davemaoite. In subducted MORB, davemaoite is expected to host roughly 5–10 mol% Ti (Ricolleau et al., 2010). Experimental work shows that Ti incorporation stabilizes the distorted tetragonal phase at room temperature (Chao et al., 2024). However, high-temperature constraints are sparse; existing measurements extend only to 12 GPa (Thomson et al., 2019). This lack of high *P*–*T* data leaves unresolved how Ti-bearing compositions modify the transition boundary and thus the depths at which the ferroelastic transition of davemaoite would occur in the mantle, and consequently whether it can account for the observed seismic anomalies.

## 2. Experimental Methods

### *2.1. Synchrotron XRD experiments using LHDAC*

We used $CaSiO_3$ powder (Alfa Aesar, 89741) as a starting material for Ca-davemaoite. The $CaSi_{0.75}Ti_{0.25}O_3$ glass starting material was synthesized by mixing $CaCO_3$, $TiO_2$, and $SiO_2$ powders in a mortar, placing the mixture in a platinum crucible, preheating at 200 °C to remove moisture, decarbonating at 900 °C overnight, and then melting the mixture at 1600 °C. The melt was subsequently quenched in water to form glass. The chemical composition, determined by electron probe microanalysis (EPMA) with a field emission microprobe (JEOL JXA-8500F, University of Hawaiʻi at Mānoa) at 15–20 keV, 20–30 nA, and a ~5 µm spot size, was $Ca_{1.02}Si_{0.75}Ti_{0.25}O_{3.02}$. Sample assemblies were prepared by drilling a hole approximately one-third of the culet diameter in the centre of a pre-indented rhenium gasket. $CaSiO_3$ or $Ca_{1.02}Si_{0.75}Ti_{0.25}O_{3.02}$ glass was formed into a 10 µm-thick pellet, sandwiched between 2.5 µm-thick Au foils used as pressure markers, ~4 µm-thick 3.0 wt.% BDD films (Sumitomo Chemical Corp.) serving as laser absorbers, and KCl pressure medium (Fig. 1a). Samples were heated using a fibre laser with simultaneous double-sided heating (wavelength λ ≈ 1070 nm, beam diameter ~25 µm), and the radiative temperature at the sample centre was measured by spectro-radiometric pyrometry. Temperature was raised gradually to ~2000 K and then decreased stepwise while XRD patterns were collected at each step. The sample was held at each temperature for several minutes prior to data collection. Temperature uncertainties were ~10% at a given pressure. Pressure was determined from the lattice parameters of the Au foils using their thermal EoS (Matsui, 2010). Symmetric LHDACs were used in all experiments. X-ray diffraction data were collected at BL10XU (Hirao et al., 2020) using a monochromatic ~30 keV beam, and the patterns were fitted using PDIndexer (Seto et al., 2010). When deriving the *c/a* ratio, the diffraction patterns were refined using a tetragonal model, which prevents the fitted *c/a* from reaching exactly 1 even when the structure is cubic. We therefore normalized *c/a* to 1 at high temperatures where the fitted values remained constant before showing an abrupt drop upon cooling (Fig. S7). This approach reflects the fact that forcing a cubic structure into a tetragonal refinement yields arbitrary *c/a* values with no physical meaning. The temperature-dependent change in *c/a*, rather than its absolute value, provides the critical evidence for this subtle structural transition.

### *2.2. Synchrotron FTIR measurements in EHDAC*

Synchrotron FTIR measurements were performed on Ca-davemaoite using the same starting material as in the XRD experiments. Fine $CaSiO_3$ powder was loaded into a diamond anvil cell and sandwiched to a thickness of approximately 10 µm with KCl as the pressure-transmitting medium. Davemaoite was pre-synthesized at 30 GPa by double-sided laser heating using a fibre laser (wavelength λ ≈ 1070 nm, beam diameter ~25 µm) at APS beamline 13-BMC (Fig. S3). The radiative temperature at the sample centre was measured with a spectro-radiometric pyrometer and reached approximately 2000 K. Successful synthesis of davemaoite was confirmed *in situ* by synchrotron XRD before FTIR measurements. Subsequent high *P–T* FTIR experiments were conducted at the Frontier infrared spectroscopy beamline 22-IR-1 of the National Synchrotron Light Source II (NSLS-II) at Brookhaven National Laboratory. Mid-IR spectra were collected using a Bruker Vertex 80 FT-IR spectrometer. Reference spectra were acquired before each measurement by probing the KCl pressure medium (Fig. S3). An EHDAC technique was employed using a tungsten wire resistive heater to generate high *P–T* (Wang et al., 2023). The DAC was enclosed in a sealed chamber with KBr windows for infrared transmission and was continuously purged with flowing nitrogen gas during heating to prevent oxidation of the diamond anvils. FTIR spectra were collected first at 300 K, followed by stepwise heating to 400 K and then in increments of 200 K up to 1000 K. Further heating was not possible due to diamond failure at higher temperatures. Temperature was measured using a K-type thermocouple placed approximately 200 µm from the sample chamber, following the configuration of Wang et al. (2023). Pressure was determined from the Raman shift of the diamond anvil culet (Akahama and Kawamura, 2006). To correct for temperature-induced Raman shifts, spectra were also collected from the back side of the anvil to quantify the thermal contribution of the peak shift at effectively ambient pressure. This thermal shift was then subtracted from the culet-side spectra to estimate pressure under high *P–T* conditions. FTIR spectra were collected only from the laser-heated region where davemaoite had been synthesized (Fig. S3). The measurement location was carefully verified before and after each acquisition using optical microscopy to ensure that all spectra were obtained from the pre-synthesized davemaoite domain.

### *2.3. Finite element method simulation*

The sample temperature was calculated by a finite element method using a software package COMSOL Multiphysics (COMSOL Inc.). Temperature distributions were calculated based on the measured laser spot size and sample thickness, using a 3D model. A homogeneous circular area with a diameter of the laser spot size was considered on both sample surfaces, and the applied heat was conducted to the sample and surrounding 10 μm-thick KCl layers. The back surface of the diamond anvils in contact with air and WC seats was set at room temperature. The thermal conductivities of davemaoite (Zhang et al., 2021) and KCl (Andersson, 1985) were taken from literature. Tetrahedral meshes with a resolution of 0.1 μm were applied. The simulated sample temperature was obtained by averaging the three-dimensional temperature distribution over a cylindrical volume corresponding to the X-ray path (3 μm in diameter) between the BDD layers (Fig. 1c).

**3. Results**

We conducted synchrotron X-ray diffraction (XRD) experiments on Ca-pure and Ti-bearing davemaoite using an LHDAC. To precisely determine the temperature-driven phase boundary, we employed a cell configuration optimized for highly stable spatiotemporal heating. A $CaSiO_3$ powder pellet (10 μm-thick) was loaded with a gold pressure marker (2.5 μm-thick), sandwiched between a boron-doped diamond (BDD) laser absorber (4 μm-thick) and a KCl pressure medium (~10 μm-thick) (Fig. 1a). This sandwich minimizes axial temperature gradients in an LHDAC (Sinmyo and Hirose, 2010), while the high thermal conductivity and strong laser-absorption of BDD reduce radial gradients and allow efficient heating at lower laser power (Fig. 1b–f). All high *P*–*T* XRD measurements were performed at the BL10XU beamline of SPring-8 (see Methods). We performed two runs for each composition (Table S1). In each run, we first compressed to the pressure of interest, and then heated at the maximum temperature for a few minutes to synthesize davemaoite. We gradually decreased the temperature to ~1200–1300 K in 0.5 W laser-power steps while collecting XRD patterns after maintaining each temperature for 30 seconds. The sample was subsequently quenched at ~1200 K, below which reliable temperatures could no longer be obtained by spectroradiometric pyrometry. Subsequently, we increased the pressure and repeated the same procedure. Note that we did not heat the sample above the melting temperature of gold pressure marker (Fig. S1). The collected XRD patterns showed only davemaoite peaks together with KCl pressure medium, gold pressure marker, and BDD laser absorber (Fig. 2a,e). For Ca-davemaoite, the (200) peak reflection of davemaoite appeared as a single peak in all datasets collected above room temperature. This observation agrees with previous experimental reports showing that the (200) doublet merges into a single peak above ~600 K (Fig. 2b). The davemaoite peaks were first fitted using a cubic structure (*Pm-3m*), and obtained the full width at half maximum (FWHM) of the (200) peak. The FWHM did not show a strong temperature dependence with decreasing temperature for data collected below ~40 GPa; however, the data collected above ~50 GPa exhibited a discontinuous increase at a specific temperature (Fig. 2c). We also fitted the davemaoite peaks with a tetragonal structure (*P4/mmm*) and obtained the *c*/*a* ratio. At high temperatures for data collected above ~50 GPa, the *c*/*a* ratio remained constant. However, at a certain temperature, the *c*/*a* ratio decreased abruptly, coinciding with the temperature at which the FWHM of the (200) peak showed a sudden increase when fitted with the cubic model (Fig. 2d). These temperatures increased systematically with pressure from 1470 K at 50 GPa to 1990 K at 84 GPa (Fig. 3).

For Ti-bearing $CaSiO_3$ davemaoite experiments, we observed a similarly sudden increase in the FWHM of (200) reflection and a corresponding sudden drop in the *c*/*a* ratio when the data were fitted using the cubic (*Pm-3m*) and tetragonal (*P4/mmm*) structures, respectively, in all runs except the highest pressure run at 62–73 GPa (Fig. 2g,h). These temperatures for Ti-bearing davemaoite also increased with pressure (Fig. 3) but were systematically higher than those of Ca-davemaoite, rising from 1300 K at 15 GPa to 2410 K at 68 GPa, that is, ~600–700 K higher at the same pressures. At room temperature, distinguishing a single versus double (200) peak in Ti-bearing davemaoite was difficult (Fig. 2f), consistent with earlier observations (Chao et al., 2024).

Both $CaSiO_3$ and $CaSi_{0.75}Ti_{0.25}O_3$ davemaoite at room temperature were identified as having the *P4/mmm* tetragonal structure, in agreement with recent studies (Chao et al., 2024). We did not observe the *I4/mcm* tetragonal structure reported in an earlier work for 17% Ti-bearing davemaoite (Chao et al., 2024), nor did we detect an orthorhombic phase, which has been documented in Al-bearing davemaoite (Kurashina et al., 2004).

## 4. Discussion

### *4.1. Tetragonal-to-cubic transition boundary of davemaoite*

The abrupt deviation of the *c*/*a* ratio from 1, together with the sudden increase in the FWHM of the (200) reflection, indicates the occurrence of the tetragonal-to-cubic phase transition in our sample. We also confirmed from synchrotron Fourier-transform infrared (FTIR) experiments that the IR spectra of Ca-davemaoite change dramatically at temperatures consistent with our observations (Fig. 3 and Fig. S3). Further support for the transition boundary comes from separate electrical conductivity measurements of $CaSiO_3$ davemaoite, which show a distinct change in temperature dependence at *P*–*T* conditions closely matching the tetragonal-to-cubic transition boundary determined here (Okuda et al., 2026). The consistency among these structural, spectroscopic, and transport signatures strongly supports the location of the transition boundary. The phase boundary of Ca-davemaoite determined in this study is approximately 500–1000 K higher than that reported in previous EHDAC studies (Komabayashi et al., 2007; Kurashina et al., 2004), yet it aligns well with predictions from theoretical calculations (Adams and Oganov, 2006; Stixrude et al., 2007, 1996; Thomson et al., 2019; Zhang et al., 2025). This discrepancy with earlier EHDAC experiments likely reflects the limited temperature ranges explored in those studies, capped at about 800 K at 30–70 GPa.

We did not observe the phase transition of Ca-davemaoite in runs conducted below 40 GPa (Fig. 3), likely because the experimental temperatures remained above the expected transition point at such pressures. Extrapolating our phase boundary to lower pressures indicates a transition temperature of roughly 500 K at ~12 GPa, consistent with previous observations at this pressure (Fig. 3; Thomson et al., 2019).

$CaSi_{0.75}Ti_{0.25}O_3$ davemaoite shows a markedly higher transition boundary, ~600–700 K above that of Ca-davemaoite, in agreement with previous reports that Ti incorporation stabilizes the tetragonal structure (Thomson et al., 2019). Our phase boundary is also compatible with that observed at 12 GPa in large volume press experiments (Thomson et al., 2019).

### *4.2. Thermal equation of state and density of davemaoite*

We fitted the high *P*–*T* data using a third-order Birch-Murnaghan equation of state, incorporating the Mie-Grüneisen-Debye (MGD) model to account for thermal pressure. For cubic Ca-davemaoite, the fitted parameters yielded $V_0$ = 45.54(16) $Å^3$ (molar volume 27.43(10) $cm^3$/mol), $K_0$ = 243(4) GPa, and $K'$ = 4.05(16), and thermoelastic parameters $\gamma_0$ = 1.47(3) and $q$ = 1.35(12) (Table S2). Debye temperature was fixed to 1100 K (Kawai and Tsuchiya, 2014; Wang et al., 1996). The obtained $V_0$, $K_0$, and $K'$ are consistent with previous *P*–*V*–*T* studies (Noguchi et al., 2013; Shim et al., 2000; Wang et al., 1996). Our $\gamma_0$ is lower than the value reported by Shim et al. (2000) [1.92(5)] and Ishii et al. (2026) [1.772], whereas our $q$ is higher than their respective estimates [0.6(3) and 1.0] (Table S2). Overall, our thermoelastic parameters are closer to those of $MgSiO_3$ perovskite (Fiquet et al., 2000).

We also fitted the data collected in the tetragonal stability field, fixing $K'$ =4. The resulting parameters were $V_0$ = 45.49(70) $Å^3$, $K_0$ = 248(6) GPa, $\gamma_0$ = 1.22(20), and $q$ = 1.08(83) (Table S2). Although uncertainties are larger due to the more limited dataset available for the tetragonal phase, the thermoelastic parameters of the cubic and tetragonal structures overlap within error. The *c*/*a* ratio of Ca-davemaoite at 300 K decreased monotonically with pressure, in excellent agreement with previous studies (Ono et al., 2004) (Fig. S4).

For $CaSi_{0.75}Ti_{0.25}O_3$ davemaoite, the EoS fit yielded $V_0$ = 48.9(2) $Å^3$ (molar volume 29.45(11) $cm^3$/mol), $K_0$ = 214(7) GPa, and $K'$ = 4.35(69), with thermoelastic parameters $\gamma_0$ = 1.21(3) and $q$ = 1.20(72) (Table S3). The Debye temperature was fixed to 900 K, based on the linear extrapolation between $CaTiO_3$ and $CaSiO_3$ (Shim et al., 2000; Woodfield et al., 1999). The obtained $V_0$, $K_0$, and $K'$ were consistent with previous room temperature EoS results for $CaSi_{0.75}Ti_{0.25}O_3$ davemaoite (Chao et al., 2024). As with Ca-davemaoite, the *c*/*a* ratio at 300 K decreased monotonically with pressure, indicating increasing tetragonal distortion at higher pressures (Fig. S4).

Using the thermal EoS derived in this study, we calculated the high *P*–*T* density of davemaoite. The room-temperature density of Ca-davemaoite agrees well with previous theoretical and experimental results (Shim et al., 2000; Sun et al., 2014), and $CaSi_{0.75}Ti_{0.25}O_3$ davemaoite shows similarly good agreement with earlier measurements at high pressure and 300 K (Chao et al., 2024) (Fig. S5). We further calculated the density of $CaSiO_3$ and $CaSi_{0.75}Ti_{0.25}O_3$ davemaoite along a slab geotherm (Syracuse et al., 2010). We found that 25% Ti incorporation reduces davemaoite density by 1.2–1.5% in the mantle. The densities of $CaSiO_3$ and $CaSi_{0.75}Ti_{0.25}O_3$ davemaoite were 2.7–3.4% and 1.5–1.9% higher than that of the normal mantle, respectively (Fig. S5). Among minerals in the MORB assemblage, bridgmanite has been suggested to be the densest phase, whereas $SiO_2$ or CF phases are among the least dense (Hirose et al., 2005; Ono et al., 2005). The estimated density of Ca-davemaoite in subducted MORB is ~3% higher than that of the MORB aggregate reported by Ono et al. (2005) but is similar to that reported in Hirose et al. (2005). Because 25 mol% Ti incorporation reduces the density of davemaoite by only 1.2–1.5%, whereas MORB bridgmanite and dense $SiO_2$ differ from the MORB aggregate by approximately +2.4–3.9% and −3.1 to −3.3%, respectively, davemaoite likely has a relatively minor effect on the bulk density of subducted MORB compared with these phases (Fig. S6).

### *4.3. Crystal structure of davemaoite and its role in lower mantle heterogeneities*

Seismological observations reveal pronounced seismic scattering in the mid-lower mantle beneath subduction zones, spanning approximately 800 to 1900 km depth (Feng et al., 2021; Li and Yuen, 2014; Niu, 2014; Yu et al., 2025). Subducted MORB is seismically slower than the surrounding mantle throughout the lower mantle and has therefore been proposed as a primary source of low-velocity anomalies (Gréaux et al., 2019; Thomson et al., 2019; Zhou et al., 2025). Indeed, models incorporating 20–60 vol% MORB can reproduce the observed seismic scatterers at depths of ~1000–1300 km, where the ferroelastic transition of dense $SiO_2$ causes MORB to become seismically slower than the surrounding mantle by approximately 7–13% (Zhou et al., 2025). At depths outside this ferroelastic transition, however, MORB produces only modest shear-wave velocity reductions of approximately 5% relative to surrounding mantle (Zhou et al., 2025), whereas seismic scatterers are associated with velocity reductions of up to ~12% (Feng et al., 2021; Kaneshima, 2016; Li and Yuen, 2014; Niu, 2014). This mismatch complicates attempts to explain the full depth extent of observed seismic scattering solely by the presence of MORB and the ferroelastic transition in dense $SiO_2$. Variations in Al content of $SiO_2$ (0–7 mol%) have been proposed as a possible explanation (Yu et al., 2025). Our results indicate that Ca-davemaoite in subducted MORB undergoes a ferroelastic transition over a broad depth range controlled by slab temperature (Fig. 3). Along cold slab geotherms, the transition occurs at depths of ~1000 km, whereas along warmer slab geotherms it is shifted to depths exceeding 2000 km. In contrast to the ferroelastic transition in dense $SiO_2$ which is primarily pressure driven, the davemaoite ferroelastic transition is strongly temperature controlled, allowing its depth to vary with slab thermal structure. This temperature sensitivity enables the generation of strong seismic scatterers over a broad depth range and provides a unified explanation for seismic scattering throughout much of the lower mantle beneath subduction zones.

We also find that Ti incorporation stabilizes the tetragonal structure. $CaSi_{0.75}Ti_{0.25}O_3$ remains tetragonal along slab geotherms, indicating that slab temperatures are insufficient to drive the transition. Assuming a linear dependence of transition temperature on Ti content, we estimate that compositions containing 5–10 mol% Ti, which is a realistic range for subducted MORB (Ricolleau et al., 2010), undergo the transition at approximately 700 to 1200 km depth along a slab geotherm (Fig. S1). These phase relations show that variations in Ti concentration within subducted MORB, combined with differences in slab geotherms, provide a mineralogical mechanism capable of explaining the observed depth variability in seismic scattering throughout the lower mantle (Feng et al., 2021; Li and Yuen, 2014; Niu, 2014; Yu et al., 2025).

Another important implication is that davemaoite within basaltic domains, such as the thick MORB piles hypothesized at the base of the mantle within the LLSVPs, may undergo a ferroelastic transition near the CMB. Geotherms at the CMB typically reach ~3700 K (Tateno et al., 2009). Our extrapolated phase diagram indicates that Ca-davemaoite can transform at the top of the thermal boundary layer, whereas Ti-bearing davemaoite with 25 mol% Ti transforms within that boundary layer (Fig. 3). More realistic Ti contents of 5–10 mol% (Ricolleau et al., 2010) would produce two transitions: an initial cubic-to-tetragonal transition at ~500–1000 km above the CMB, followed by a return to the cubic structure within the thermal boundary layer (Fig. S1). Given that the LLSVPs occupy the region from the CMB to ~1000 km above it (Garnero et al., 2016), the transition depths derived from our phase diagram indicate that MORB-derived davemaoite likely contributes to the LLSVP seismic signature. Our finding aligns with the view that the LLSVPs are enriched in basaltic materials, supported by mantle convection simulations (Nakagawa and Tackley, 2004) and electrical conductivity measurements (Ohta et al., 2010). Variations in Ti content in MORB-derived davemaoite therefore provide a coherent explanation for both seismic scattering in the mid-lower mantle and the anomalous seismic signatures of the LLSVPs, underscoring its role in generating mantle heterogeneity.

**CRediT authorship contribution statement**

**Yoshiyuki Okuda:** Conceptualization, Methodology, Investigation, Writing – original draft, Writing – review & editing. **Bin Chen:** Conceptualization, Methodology, Investigation, Writing – original draft, Writing – review & editing. **Juliana Peckenpaugh:** Investigation, Writing – review & editing. **S. Kawaguchi-Imada:** Investigation, Resources, Writing – review & editing. **Hirokazu Kadobayashi:** Investigation, Resources, Writing – review & editing. **Z. Liu:** Investigation, Resources, Writing – review & editing. **D. Zhang:** Investigation, Writing – review & editing. **K.-H. Chao**: Investigation, Writing – review & editing.

**Declaration of competing interests**

The authors declare that they have no competing interests.

**Data and materials availability**

All data needed to evaluate the conclusions in the paper are available in the main text and/or the Supplementary Materials. Additional data related to this paper may be requested from the authors.

**Acknowledgements**

We thank Prof. Kenji Ohta for his assistance with the COMSOL analysis. This work was supported by the JSPS KAKENHI Grants No. 22J00928 (Y.O.), 22K21344 (Y.O.), and by the U.S. National Science Foundation grant EAR-2127807 (B.C.) and NASA grant 80NSSC22K0138 (B.C.). The synchrotron X-ray diffraction experiments were performed at the BL10XU, SPring-8 (Proposal No. 2024B1457), and the FTIR measurements were conducted at the 22-IR-1, NSLS-II (Proposal No. 317554).

**Supplementary materials**

Figs. S1–S7; Tables S1–S3.

## References

Adams, D.J., Oganov, A.R., 2006. *Ab initio* molecular dynamics study of $CaSiO_3$ perovskite at P−T conditions of Earth's lower mantle. Phys. Rev. B 73, 184106. https://doi.org/10.1103/PhysRevB.73.184106

Akahama, Y., Kawamura, H., 2006. Pressure calibration of diamond anvil Raman gauge to 310GPa. J. Appl. Phys. 5.

Andersson, P., 1985. Thermal conductivity under pressure and through phase transitions in solid alkali halides. I. Experimental results for KCl, KBr, KI, RbCl, RbBr and RbI. J. Phys. C: Solid State Phys. 18, 3943–3955. https://doi.org/10.1088/0022-3719/18/20/020

Chao, K.-H., Berrada, M., Wang, S., Peckenpaugh, J., Zhang, D., Chariton, S., Prakapenka, V., Chen, B., 2024. Structure and equation of state of Ti-bearing davemaoite: new insights into the chemical heterogeneity in the lower mantle. American Mineralogist. https://doi.org/10.2138/am-2023-9104

Chen, H., Shim, S.-H., Leinenweber, K., Prakapenka, V., Meng, Y., Prescher, C., 2018. Crystal structure of $CaSiO_3$ perovskite at 28–62 GPa and 300 K under quasi-hydrostatic stress conditions. American Mineralogist 103, 462–468. https://doi.org/10.2138/am-2018-6087

Feng, J., Yao, H., Wang, Y., Poli, P., Mao, Z., 2021. Segregated oceanic crust trapped at the bottom mantle transition zone revealed from ambient noise interferometry. Nat. Commun. 12, 2531. https://doi.org/10.1038/s41467-021-22853-2

Fiquet, G., Dewaele, A., Andrault, D., Kunz, M., Le Bihan, T., 2000. Thermoelastic properties and crystal structure of $MgSiO_3$ perovskite at lower mantle pressure and temperature conditions. Geophys. Res. Lett. 27, 21–24. https://doi.org/10.1029/1999GL008397

Garnero, E.J., McNamara, A.K., Shim, S.-H., 2016. Continent-sized anomalous zones with low seismic velocity at the base of Earth's mantle. Nature Geosci 9, 481–489. https://doi.org/10.1038/ngeo2733

Gréaux, S., Irifune, T., Higo, Y., Tange, Y., Arimoto, T., Liu, Z., Yamada, A., 2019. Sound velocity of $CaSiO_3$ perovskite suggests the presence of basaltic crust in the Earth's lower mantle. Nature 565, 218–221. https://doi.org/10.1038/s41586-018-0816-5

Hirao, N., Kawaguchi, S.I., Hirose, K., Shimizu, K., Ohtani, E., Ohishi, Y., 2020. New developments in high-pressure X-ray diffraction beamline for diamond anvil cell at SPring-8. Matter Radiat. Extrem. 5, 018403. https://doi.org/10.1063/1.5126038

Hirose, K., Takafuji, N., Sata, N., Ohishi, Y., 2005. Phase transition and density of subducted MORB crust in the lower mantle. Earth Planet. Sci. Lett. 237, 239–251. https://doi.org/10.1016/j.epsl.2005.06.035

Inada, M., Okuda, Y., Oka, K., Kuwahara, H., Gréaux, S., Hirose, K., 2025. Electrical Conductivity of Hydrous $SiO_2$ : Implications for the Superionic State and High Conductivity Anomalies Beneath Subduction Zones. JGR Solid Earth 130, e2025JB032641. https://doi.org/10.1029/2025JB032641

Irifune, T., Ringwood, A.E., 1993. Phase transformations in subducted oceanic crust and buoyancy relationships at depths of 600–800 km in the mantle. Earth and Planetary Science Letters 117, 101–110. https://doi.org/10.1016/0012-821X(93)90120-X

Ishii, T., Takaichi, G., Nishihara, Y., Matsukage, K.N., Itoh, S., Lin, Y., Irifune, T., Ikuta, D., Zhao, B., Diyalanthonige, D.H.F., Tsujino, N., Kakizawa, S., Higo, Y., 2026. Limited water incorporation in davemaoite under lower-mantle conditions. Commun Earth Environ. https://doi.org/10.1038/s43247-026-03856-7

Kaneshima, S., 2016. Seismic scatterers in the mid-lower mantle. Physics of the Earth and Planetary Interiors 257, 105–114. https://doi.org/10.1016/j.pepi.2016.05.004

Katsura, T., 2022. A Revised Adiabatic Temperature Profile for the Mantle. JGR Solid Earth 127, e2021JB023562. https://doi.org/10.1029/2021JB023562

Kawai, K., Tsuchiya, T., 2014. *P-V-T* equation of state of cubic $CaSiO_3$ perovskite from first-principles computation. JGR Solid Earth 119, 2801–2809. https://doi.org/10.1002/2013JB010905

Komabayashi, T., Hirose, K., Sata, N., Ohishi, Y., Dubrovinsky, L.S., 2007. Phase transition in $CaSiO_3$ perovskite. Earth Planet. Sci. Lett. 260, 564–569. https://doi.org/10.1016/j.epsl.2007.06.015

Kudo, Y., Hirose, K., Murakami, M., Asahara, Y., Ozawa, H., Ohishi, Y., Hirao, N., 2012. Sound velocity measurements of $CaSiO_3$ perovskite to 133GPa and implications for lowermost mantle seismic anomalies. Earth Planet. Sci. Lett. 349–350, 1–7. https://doi.org/10.1016/j.epsl.2012.06.040

Kurashina, T., Hirose, K., Ono, S., Sata, N., Ohishi, Y., 2004. Phase transition in Al-bearing $CaSiO_3$ perovskite: implications for seismic discontinuities in the lower mantle. Physics of the Earth and Planetary Interiors 145, 67–74. https://doi.org/10.1016/j.pepi.2004.02.005

Lai, X., Zhu, F., Gao, J., Greenberg, E., Prakapenka, V.B., Meng, Y., Chen, B., 2022. Melting of the Fe-C-H System and Earth's Deep Carbon-Hydrogen Cycle. Geophys. Res. Lett. 49. https://doi.org/10.1029/2022GL098919

Li, J., Yuen, D.A., 2014. Mid-mantle heterogeneities associated with izanagi plate: implications for regional mantle viscosity. Earth Planet. Sci. Lett. 385, 137–144. https://doi.org/10.1016/j.epsl.2013.10.042

Matsui, M., 2010. High temperature and high pressure equation of state of gold. J. Phys.: Conf. Ser. 215, 012197. https://doi.org/10.1088/1742-6596/215/1/012197

Nakagawa, T., Tackley, P.J., 2004. Thermo-chemical structure in the mantle arising from a three-component convective system and implications for geochemistry. Phys. Earth Planet. Inter. 146, 125–138. https://doi.org/10.1016/j.pepi.2003.05.006

Niu, F., 2014. Distinct compositional thin layers at mid-mantle depths beneath northeast China revealed by the USArray. Earth Planet. Sci. Lett. 402, 305–312. https://doi.org/10.1016/j.epsl.2013.02.015

Noguchi, M., Komabayashi, T., Hirose, K., Ohishi, Y., 2013. High-temperature compression experiments of $CaSiO_3$ perovskite to lowermost mantle conditions and its thermal equation of state. Phys. Chem. Minerals 40, 81–91. https://doi.org/10.1007/s00269-012-0549-1

Ohta, K., Hirose, K., Ichiki, M., Shimizu, K., Sata, N., Ohishi, Y., 2010. Electrical conductivities of pyrolitic mantle and MORB materials up to the lowermost mantle conditions. Earth Planet. Sci. Lett. 289, 497–502. https://doi.org/10.1016/j.epsl.2009.11.042

Okuda, Y., Chen, B., Peckenpaugh, J., Chao, K.-H., Kadobayashi, H., Liu, Z., Zhang, D., 2026. The tetragonal-cubic transition of davemaoite: Implications to lower mantle seismic anomalies. Earth Planet. Sci. Lett. submitted.

Okuda, Y., Kimura, S., Ohta, K., Park, Y., Wakamatsu, T., Mashino, I., Hirose, K., 2021. A cylindrical SiC heater for an externally heated diamond anvil cell to 1500 K. Rev. Sci. Instrum. 92, 015119. https://doi.org/10.1063/5.0036551

Ono, S., Ohishi, Y., Isshiki, M., Watanuki, T., 2005. In situ X-ray observations of phase assemblages in peridotite and basalt compositions at lower mantle conditions: Implications for density of subducted oceanic plate. J. Geophys. Res. 110, 2004JB003196. https://doi.org/10.1029/2004JB003196

Ono, S., Ohishi, Y., Mibe, K., 2004. Phase transition of Ca-perovskite and stability of Al-bearing Mg-perovskite in the lower mantle. Am. Mineral. 89, 1480–1485. https://doi.org/10.2138/am-2004-1016

Ricolleau, A., Perrillat, J., Fiquet, G., Daniel, I., Matas, J., Addad, A., Menguy, N., Cardon, H., Mezouar, M., Guignot, N., 2010. Phase relations and equation of state of a natural MORB: implications for the density profile of subducted oceanic crust in the Earth's lower mantle. J. Geophys. Res. 115, 2009JB006709. https://doi.org/10.1029/2009JB006709

Sagatova, D.N., Shatskiy, A.F., Sagatov, N.E., Litasov, K.D., 2021. Phase Relations in $CaSiO_3$ System up to 100 GPa and 2500 K. Geochem. Int. 59, 791–800. https://doi.org/10.1134/s0016702921080073

Seto Y., Nishio-Hamane D., Nagai T., Sata N., 2010. Development of a Software Suite on X-ray Diffraction Experiments. The Review of High Pressure Science and Technology 20, 269–276. https://doi.org/10.4131/jshpreview.20.269

Shim, S.-H., Duffy, T.S., Shen, G., 2000. The stability and P-V-T equation of state of $CaSiO_3$ perovskite in the Earth's lower mantle. J. Geophys. Res. 105, 25955–25968. https://doi.org/10.1029/2000JB900183

Sinmyo, R., Hirose, K., 2010. The Soret diffusion in laser-heated diamond-anvil cell. Phys. Earth Planet. Inter. 180, 172–178. https://doi.org/10.1016/j.pepi.2009.10.011

Stixrude, L., Lithgow-Bertelloni, C., Kiefer, B., Fumagalli, P., 2007. Phase stability and shear softening in Ca Si O 3 perovskite at high pressure. Phys. Rev. B 75, 024108. https://doi.org/10.1103/PhysRevB.75.024108

Stixrude, L., Ronald Cohen, Yu, R., Krakauer, H., 1996. Prediction of phase transition in CaSi03 perovskite and implications for lower mantle structure. Am. Mineral. 81, 1293–1296.

Sun, N., Bian, H., Zhang, Y., Lin, J.-F., Prakapenka, V.B., Mao, Z., 2022. High-pressure experimental study of tetragonal CaSiO3-perovskite to 200 GPa. American Mineralogist 107, 110–115. https://doi.org/10.2138/am-2021-7913

Sun, T., Zhang, D.-B., Wentzcovitch, R.M., 2014. Dynamic stabilization of cubic $CaSiO_3$ perovskite at high temperatures and pressures from *ab initio* molecular dynamics. Phys. Rev. B 89, 094109. https://doi.org/10.1103/PhysRevB.89.094109

Syracuse, E.M., Van Keken, P.E., Abers, G.A., 2010. The global range of subduction zone thermal models. Phys. Earth Planet. Inter. 183, 73–90. https://doi.org/10.1016/j.pepi.2010.02.004

Tateno, S., Hirose, K., Sata, N., Ohishi, Y., 2009. Determination of post-perovskite phase transition boundary up to 4400 K and implications for thermal structure in D″ layer. Earth Planet. Sci. Lett. 277, 130–136. https://doi.org/10.1016/j.epsl.2008.10.004

Thomson, A.R., Crichton, W.A., Brodholt, J.P., Wood, I.G., Siersch, N.C., Muir, J.M.R., Dobson, D.P., Hunt, S.A., 2019. Seismic velocities of CaSiO3 perovskite can explain LLSVPs in Earth's lower mantle. Nature 572, 643–647. https://doi.org/10.1038/s41586-019-1483-x

Tschauner, O., Huang, S., Yang, S., Humayun, M., Liu, W., Gilbert Corder, S.N., Bechtel, H.A., Tischler, J., Rossman, G.R., 2021. Discovery of davemaoite, $CaSiO_3$ -perovskite, as a mineral from the lower mantle. Science 374, 891–894. https://doi.org/10.1126/science.abl8568

Wang, S., Berrada, M., Chao, K.-H., Lai, X., Zhu, F., Zhang, D., Chariton, S., Prakapenka, V.B., Sinogeikin, S., Chen, B., 2023. Externally Heated Diamond ANvil Cell Experimentation (EH-DANCE) for studying materials and processes under extreme conditions. Rev. Sci. Instrum. 94, 123902. https://doi.org/10.1063/5.0180103

Wang, Y., Weidner, D.J., Guyot, F., 1996. Thermal equation of state of $CaSiO_3$ perovskite. J. Geophys. Res. 101, 661–672. https://doi.org/10.1029/95JB03254

Woodfield, B.F., Shapiro, J.L., Stevens, R., Boerio-Goates, J., Putnam, R.L., Helean, K.B., Navrotsky, A., 1999. Molar heat capacity and thermodynamic functions for. J. Chem. Thermodyn. 31, 1573–1583. https://doi.org/10.1006/jcht.1999.0556

Wu, F., Sun, Y., Wan, T., Wu, S., Wentzcovitch, R.M., 2024. Deep-Learning-Based Prediction of the Tetragonal → Cubic Transition in Davemaoite. Geophysical Research Letters 51, e2023GL108012. https://doi.org/10.1029/2023GL108012

Yu, Y., Zhang, Youyue, Li, L., Zhang, X., Wang, D., Mao, Z., Sun, N., Zhang, Yanyao, Li, X., Li, W., Speziale, S., Zhang, D., Lin, J., Yoshino, T., 2025. Unraveling the Complex Features of the Seismic Scatterers in the Mid-Lower Mantle Through Phase Transition of (Al, H)-Bearing Stishovite. Geophysical Research Letters 52, e2024GL114146. https://doi.org/10.1029/2024GL114146

Zhang, C., Yang, J.-Y., Sun, T., Zhang, H., Brodholt, J.P., 2025. Strong precursor softening in cubic $CaSiO_3$ perovskite. Proc. Natl. Acad. Sci. U.S.A. 122, e2410910122. https://doi.org/10.1073/pnas.2410910122

Zhang, Z., Zhang, D.-B., Onga, K., Hasegawa, A., Ohta, K., Hirose, K., Wentzcovitch, R.M., 2021. Thermal conductivity of $CaSiO_3$ perovskite at lower mantle conditions. Phys. Rev. B 104, 184101. https://doi.org/10.1103/PhysRevB.104.184101

Zhou, W.-Y., Hao, M., Su, W., Kim, T., Chen, S., Shim, S.-H., Zhang, D., Nguyen, P.Q.H., Armstrong, K., Zhang, J.S., 2025. Elasticity of davemaoite as a primary contributor to lower-mantle heterogeneities. Science 390, 935–939. https://doi.org/10.1126/science.adx8356

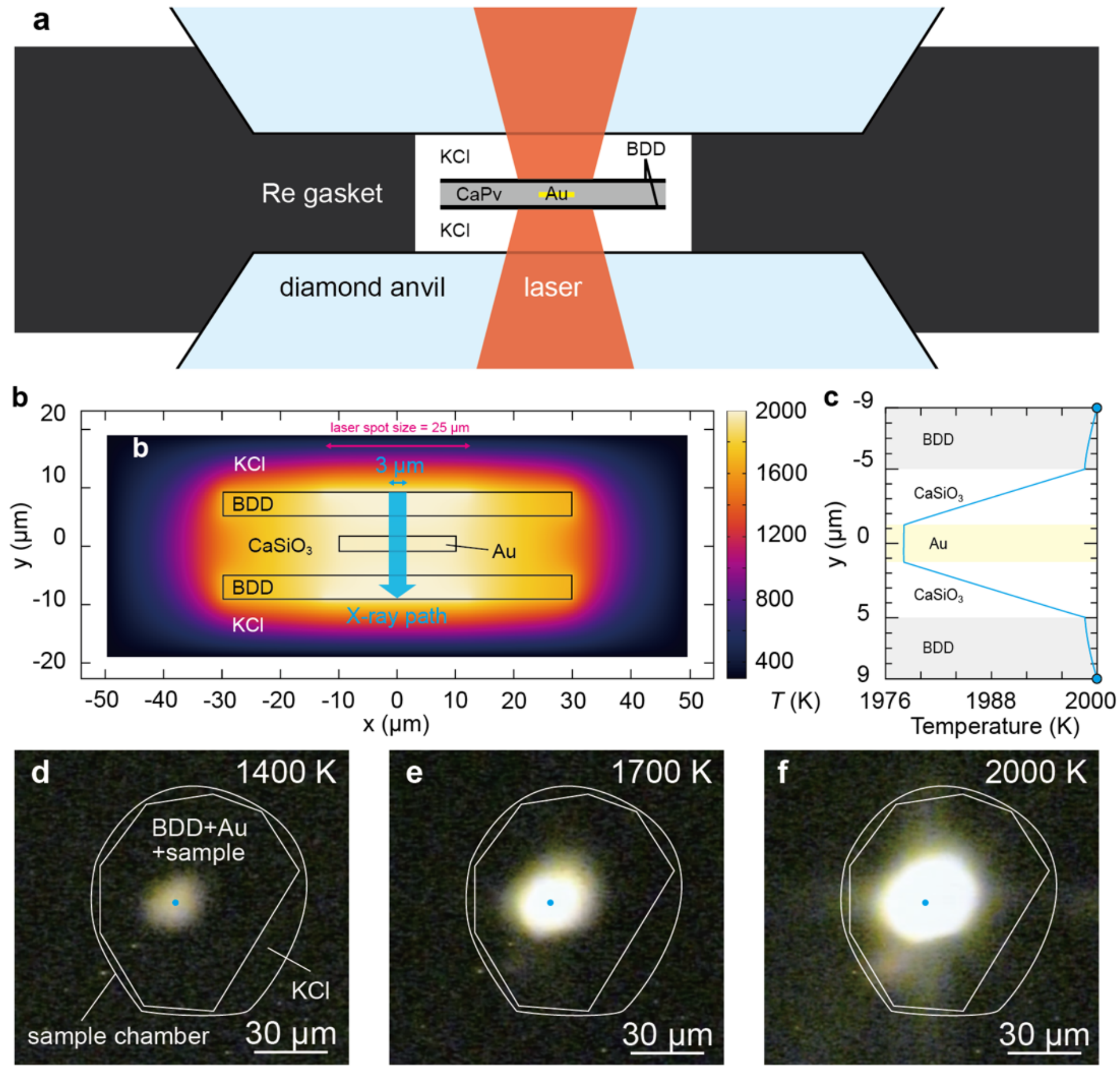

**Fig. 1**. **Experimental assembly and laser heating performance.** (a) Schematic of the cell assembly. (b) Simulated two-dimensional temperature distribution in a DAC sample when the boron-doped diamond laser absorber was heated to 2000 K. (c) One-dimensional temperature profile along the X-ray path, showing a homogeneous heating field with a minimal thermal gradient (~±1% across the X-ray path). Circles show the BDD surface temperature measured from thermal radiation. (d–f) Optical micrographs of the DAC samples heated to 1400, 1700, and 2000 K in run #1, demonstrating uniform spatial heating (see also Video S1 for temporal stability). The blue dots indicate the X-ray spot.

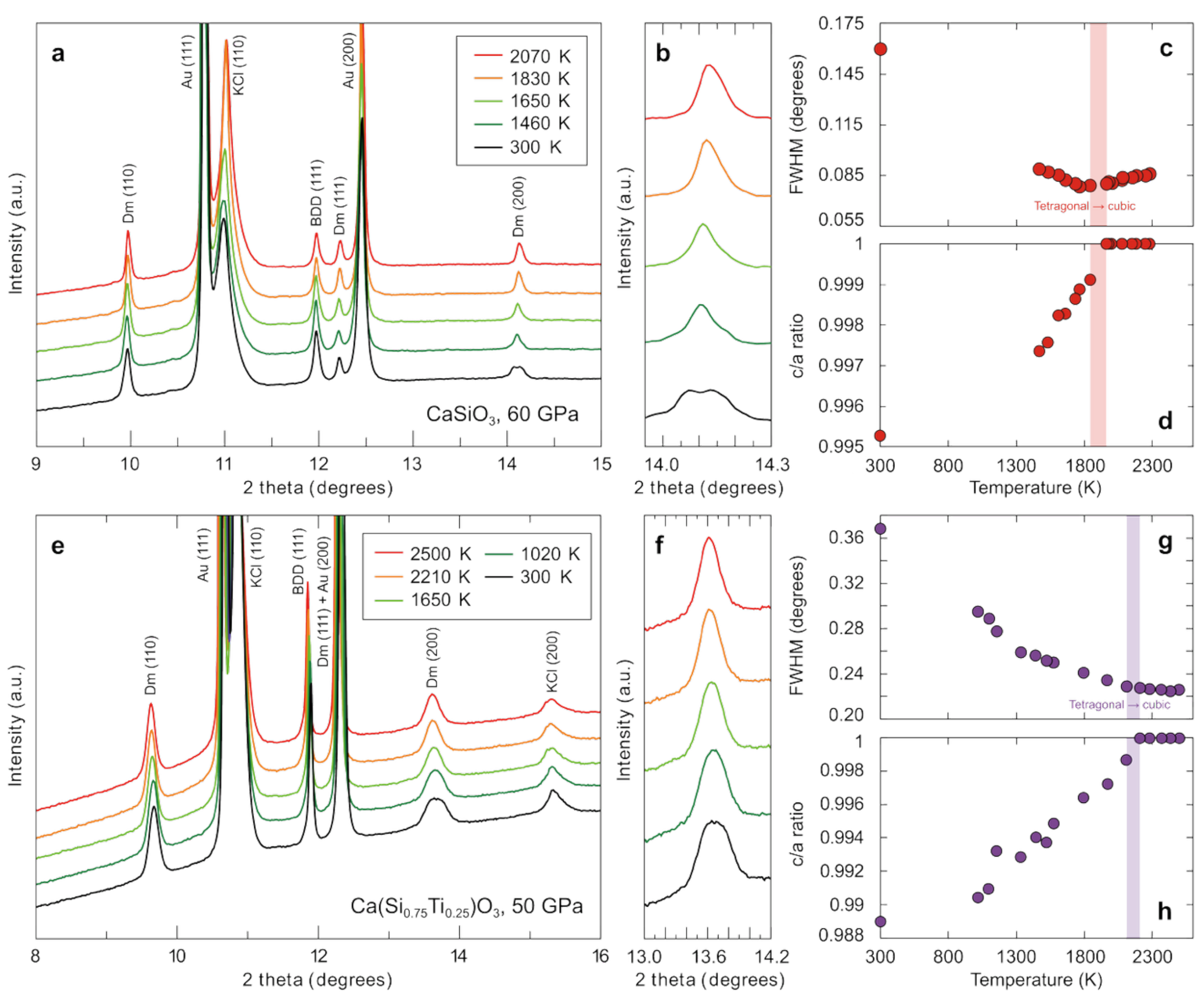

**Fig. 2. Integrated XRD patterns, FWHM of the (200) peak, and *c/a* ratio of davemaoite.** (a–d) $CaSiO_3$ davemaoite collected at ~60 GPa in run #1. (e–h) $CaSi_{0.75}Ti_{0.25}O_3$ davemaoite at ~ 50 GPa in run #1. See Fig. S2 for XRD data collected in other runs.

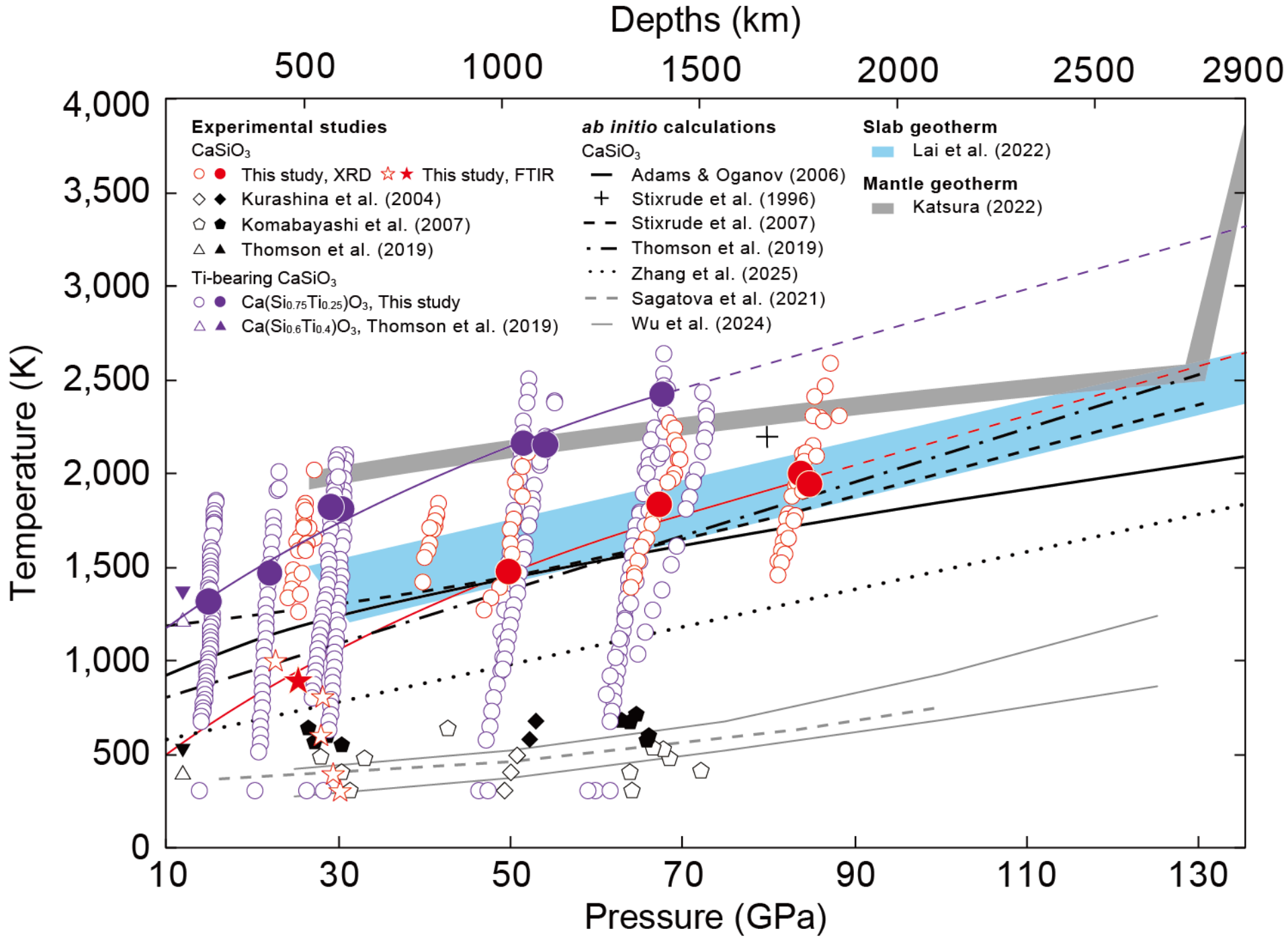

**Fig. 3. Pressure-temperature conditions of tetragonal-cubic transition.** Red and purple open circles indicate the measurement *P–T* conditions for $CaSiO_3$ and $CaSi_{0.75}Ti_{0.25}O_3$ in this study, respectively. Filled red and purple circles denote the conditions where we observed the discontinuity in both the temperature dependence of FWHM of (200) peak and *c*/*a* ratio for $CaSiO_3$, and $CaSi_{0.75}Ti_{0.25}O_3$ davemaoite, respectively. Red open stars indicate the *P–T* conditions of the synchrotron FTIR measurements on $CaSiO_3$ davemaoite performed in this study. The filled star marks the midpoint between 800 and 1000 K, across which we observed significant changes in the FTIR spectra, suggesting a phase transition. Diamond, pentagon, and triangles indicate those of $CaSiO_3$ for Kurashina et al. (2004), Komabayashi et al. (2007), and Thomson et al. (2019), respectively. Purple triangles are for $CaSi_{0.6}Ti_{0.4}O_3$ davemaoite (Thomson et al., 2019). Open and filled symbols indicate tetragonal and cubic structure, respectively. Solid, broken, dot-broken, dotted, dashed, and thin solid curves, and plus symbol indicate the tetragonal-to-cubic transition predicted by *ab initio* calculations in Adams and Oganov (2006), Stixrude et al. (2007), Thomson et al. (2019), Zhang et al. (2025), Sagatova et al. (2021), Wu et al. (2024), and Stixrude et al. (1996), respectively. Grey band

represents a normal mantle geotherm (Katsura, 2022), including a thermal boundary layer above the core-mantle boundary (CMB), assuming a CMB temperature of 3700 K (Tateno et al., 2009). Blue band denotes the subducted slab geotherm (Lai et al., 2022; Syracuse et al., 2010).

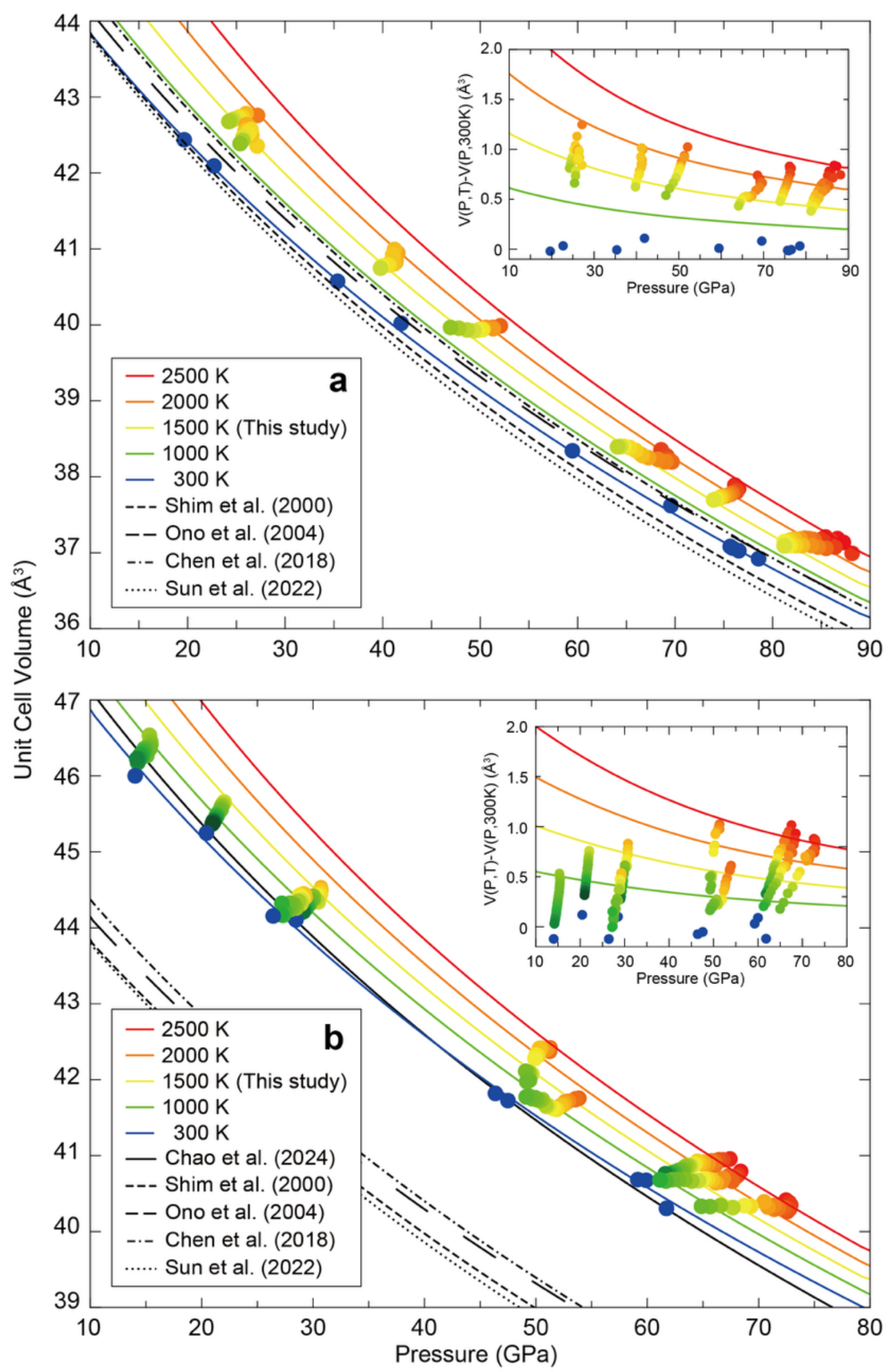

**Fig. 4. Pressure–volume–temperature relations of (a) $CaSiO_3$, and (b) $CaSi_{0.75}Ti_{0.25}O_3$ davemaoite.** Coloured symbols represent the experimental data from this study, where colour indicates temperature (blue to red for 300–2500 K). Solid coloured curves are fitted equations of state for each temperature, where $CaSiO_3$ and $CaSi_{0.75}Ti_{0.25}O_3$ are results fitted by cubic and tetragonal structures, respectively. Black lines denote previous experimental and theoretical results for comparison: solid (Chao et al., 2024), dashed (Shim et al., 2000), long-dashed (Ono et al., 2004), dash-dotted (Chen et al., 2018), and dotted (Sun et al., 2022). The inset shows the volume difference $\Delta V = V(P, T) - V(P, 300\text{ K})$ as a function of pressure.

## Supplementary information

# The tetragonal-cubic transition of davemaoite: Implications for lower mantle seismic anomalies

Yoshiyuki Okuda[1,2,*], Bin Chen[1,*], Juliana Peckenpaugh[1], Keng-Hsien Chao[1], Saori Kawaguchi-Imada[3], Hirokazu Kadobayashi[3], Zhenxian Liu[4], Dongzhou Zhang[5]

[1] *Hawai'i Institute of Geophysics and Planetology, University of Hawai'i at Manoa, Honolulu, Hawai'i 96822, USA*

[2] *Department of Earth and Planetary Sciences, Institute of Science Tokyo, Meguro, Tokyo 152-8551, Japan*

[3] *SPring-8, Japan Synchrotron Radiation Research Institute, Sayo, Hyogo 679-5198, Japan*

[4] *National Synchrotron Light Source II, Brookhaven National Laboratory, Upton, New York 11973, USA*

[5] *Center for Advanced Radiation Sources, The University of Chicago, Chicago, Illinois 60637, USA*

## Supplementary Figures

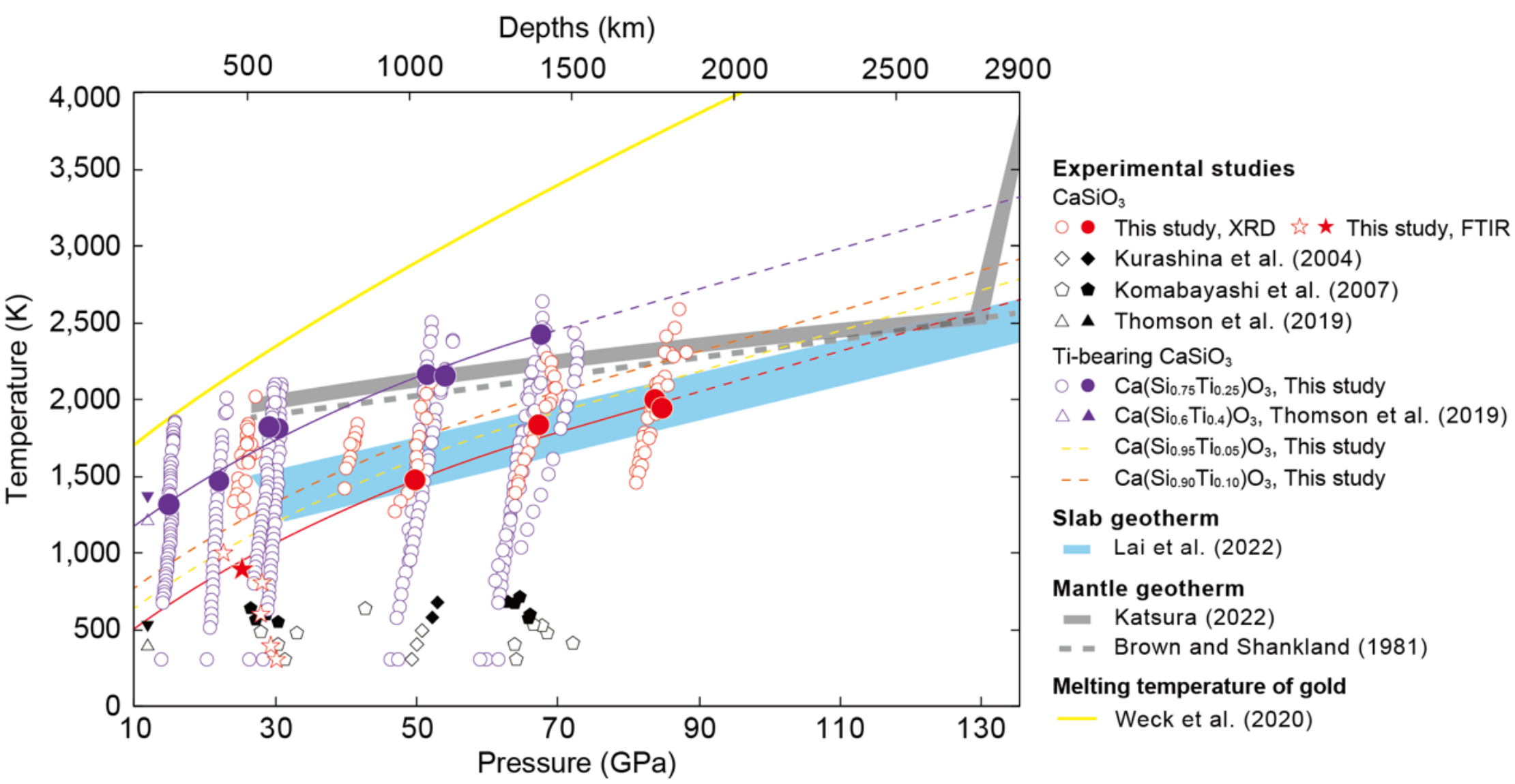

**Fig. S1. Investigated *P–T* conditions plotted together with gold melting temperature and different mantle geotherms.** Red and purple circles represent $CaSiO_3$ and $CaSi_{0.75}Ti_{0.25}O_3$ in this study. Filled red and purple circles denote the conditions where we observed discontinuities in both the FWHM of the (200) peak and the *c/a* ratio for $CaSiO_3$, and $CaSi_{0.75}Ti_{0.25}O_3$ davemaoite, respectively. Purple curve indicates the phase boundary of $CaSi_{0.75}Ti_{0.25}O_3$ determined in this study fitted with a polynomial function. The red curve indicates the phase boundary of $CaSiO_3$ davemaoite, constructed by translating the Ti-bearing boundary in parallel while assuming the same curvature. Yellow and orange curves denote the inferred boundaries for $CaSi_{0.95}Ti_{0.05}O_3$ and $CaSi_{0.90}Ti_{0.10}O_3$ davemaoite, respectively, assuming a linear dependence of transition temperature on Ti content. Diamond, pentagon, and triangles indicate those of $CaSiO_3$ for Kurashina et al. (2004), Komabayashi et al. (2007), and Thomson et al. (2019), respectively. Purple triangles are for $CaSi_{0.6}Ti_{0.4}O_3$ davemaoite (Thomson et al., 2019). Open and filled symbols indicate tetragonal and cubic structure, respectively. Solid, broken, dot-broken, dotted, dashed, and thin solid curves, and plus symbol indicate the tetragonal-to-cubic transition predicted by *ab initio* calculations in Adams and Oganov (2006), Stixrude et al. (2007), Thomson et al. (2019), Zhang et al. (2025),

Sagatova et al. (2021), Wu et al. (2024), and Stixrude et al. (1996), respectively. Red stars indicate the *P–T* conditions of the synchrotron FTIR measurements on $CaSiO_3$ davemaoite performed in this study (see Fig. S3 for the corresponding FTIR spectra). The filled star marks the midpoint between 800 and 1000 K, across which we observed significant changes in the FTIR spectra, suggesting a phase transition, which is consistent with the extrapolated phase transition boundary determined from XRD measurements. Grey band represents a normal mantle geotherm (Katsura, 2022), including a thermal boundary layer above the core-mantle boundary (CMB), assuming a CMB temperature of 3700 K (Tateno et al., 2009). Grey broken curve shows a conventional normal mantle geotherm (Brown and Shankland, 1981). Blue band denotes the subducted slab geotherm (Lai et al., 2022; Syracuse et al., 2010). Yellow curve indicates the melting temperature of gold at high *P–T* (Weck et al., 2020).

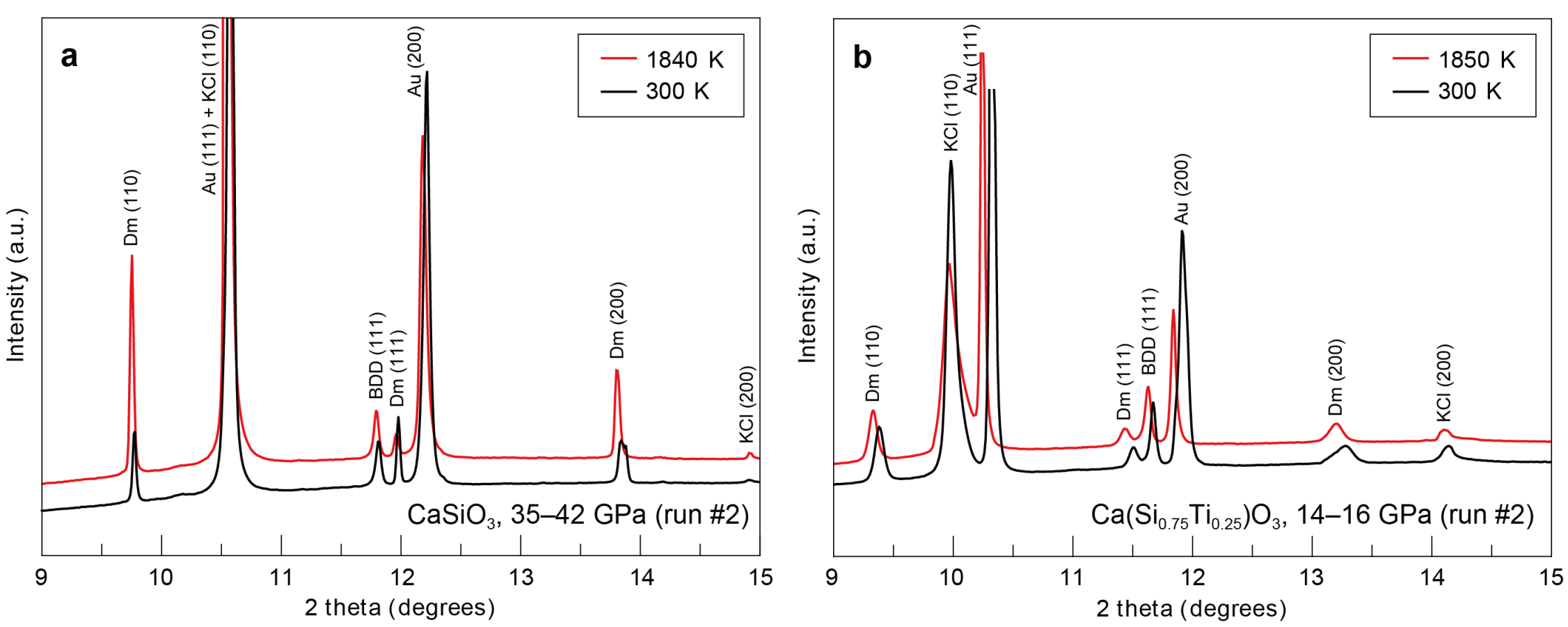

**Fig. S2. XRD patterns obtained in this study of (a) $CaSiO_3$ davemaoite collected in run #2, and (b) $CaSi_{0.75}Ti_{0.25}O_3$ davemaoite in run #2.**

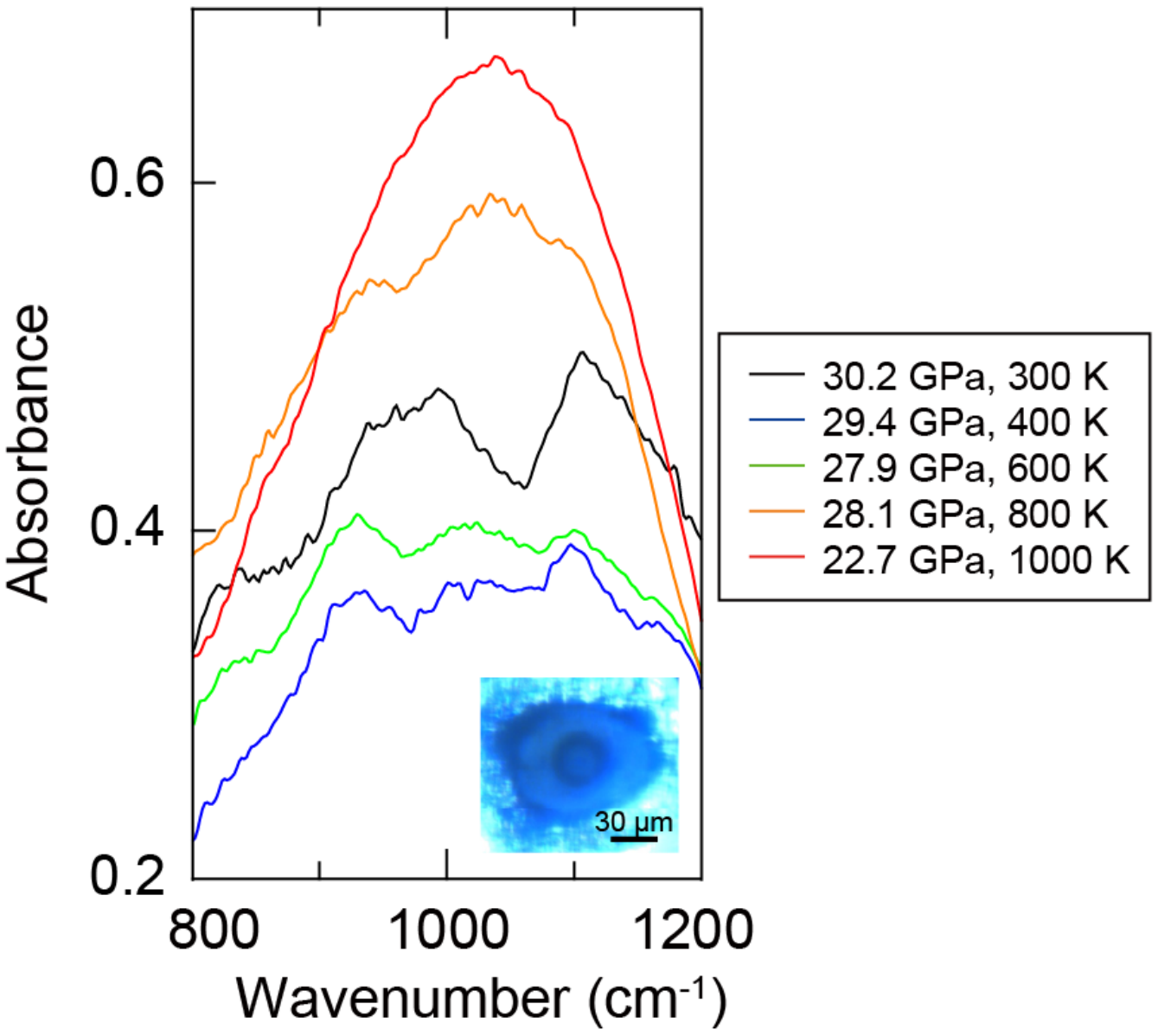

**Fig. S3. FTIR spectra of $CaSiO_3$ davemaoite collected at 23 to 30 GPa and 300 to 1000 K.** Different colours indicate different temperatures. A sharp change in absorbance and a significant change in the Si–O vibrational features around ~1000 $cm^{-1}$ are observed between 800 and 1000 K. In contrast, no clear change in the FTIR patterns is detected around ~500 K (within the 400 to 600 K range), where a transition has been reported in previous studies (Komabayashi et al., 2007). The inset shows an optical micrograph of the sample chamber during heating. The spherical domain at the centre corresponds to pre-synthesized $CaSiO_3$ davemaoite, surrounded by KCl pressure medium.

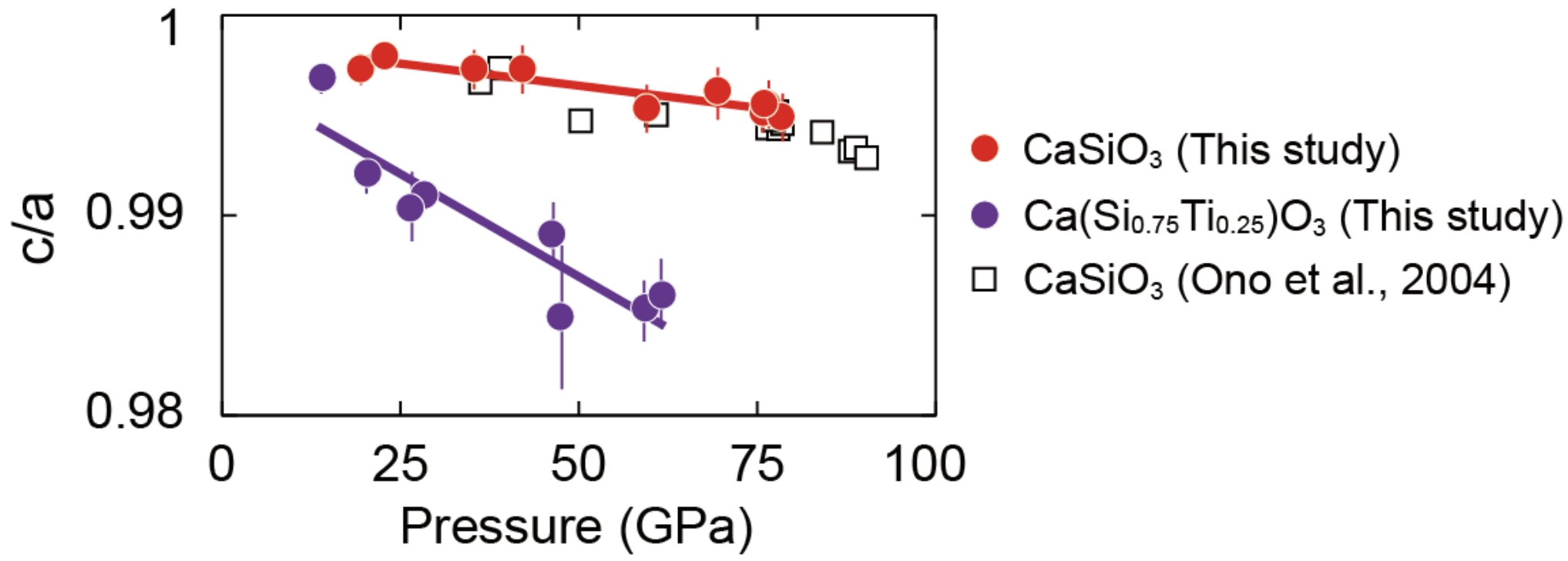

**Fig. S4. The *c*/*a* ratio of davemaoite at high pressures and room temperature.** Red and purple circles represent $CaSiO_3$ and $CaSi_{0.75}Ti_{0.25}O_3$ measured in this study. Open squares show $CaSiO_3$ reported by Ono et al. (2004).

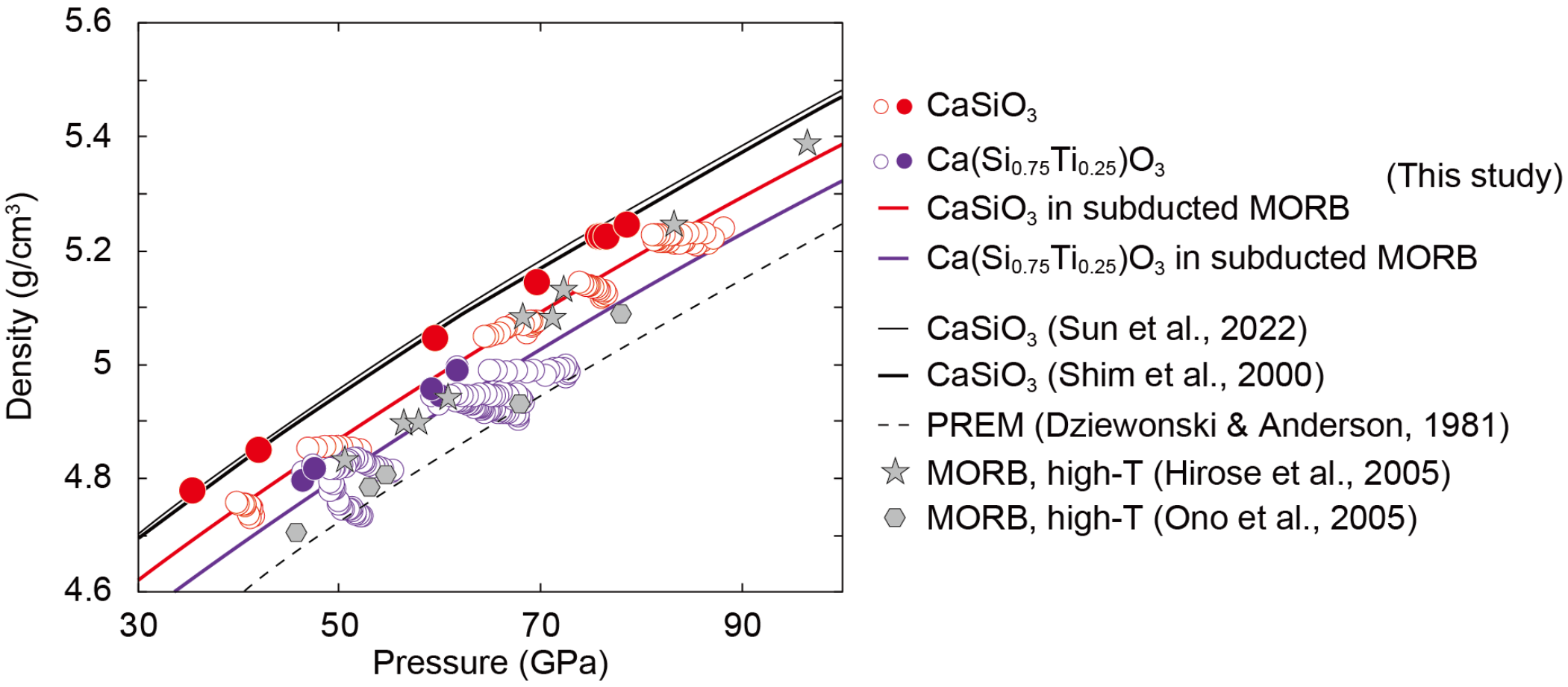

**Fig. S5. Density of davemaoite.** Red and purple circles represent $CaSiO_3$ and $CaSi_{0.75}Ti_{0.25}O_3$ measured in this study. Filled and open symbols show densities at 300 K and at high *P*–*T*, respectively. Red and purple curves denote the calculated densities of $CaSiO_3$ and $CaSi_{0.75}Ti_{0.25}O_3$ davemaoite along a cold slab geotherm. Grey (Sun et al., 2022) and black (Shim et al., 2000) curves show previously reported density profiles for $CaSiO_3$ davemaoite. The grey dashed curve represents the PREM density profile (Dziewonski and Anderson, 1981). Grey stars (Hirose et al., 2005) and hexagons (Ono et al., 2005) indicate densities of subducted MORB assemblages from earlier DAC studies.

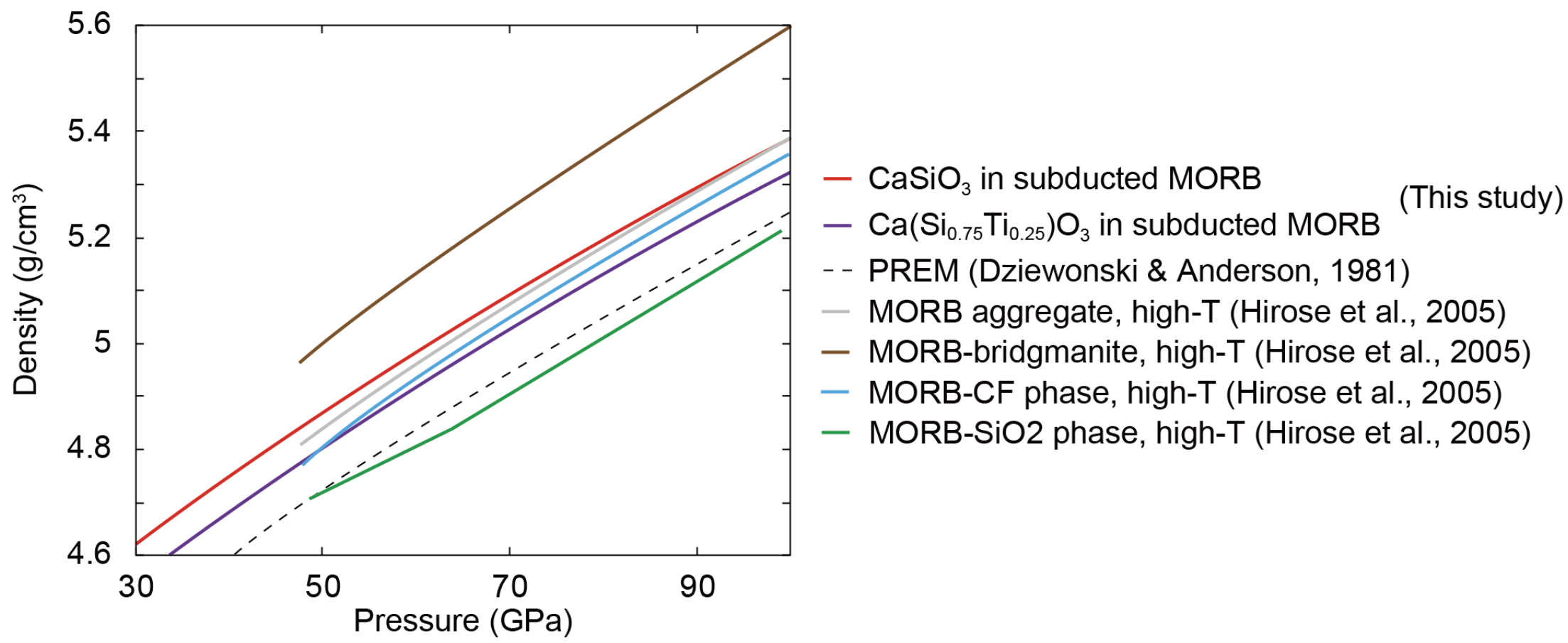

**Fig. S6. Density of davemaoite in subducted slab.** The red line shows $CaSiO_3$ in subducted MORB, and the purple line shows $CaSi_{0.75}Ti_{0.25}O_3$ in subducted MORB determined in this study. For comparison, the dashed black line indicates the Preliminary Reference Earth Model (PREM; Dziewonski and Anderson, 1981). Grey and green lines show densities of MORB aggregate, MORB bridgmanite, MORB davemaoite, and MORB $SiO_2$ phase calculated at high pressures and high temperatures (Hirose et al., 2005).

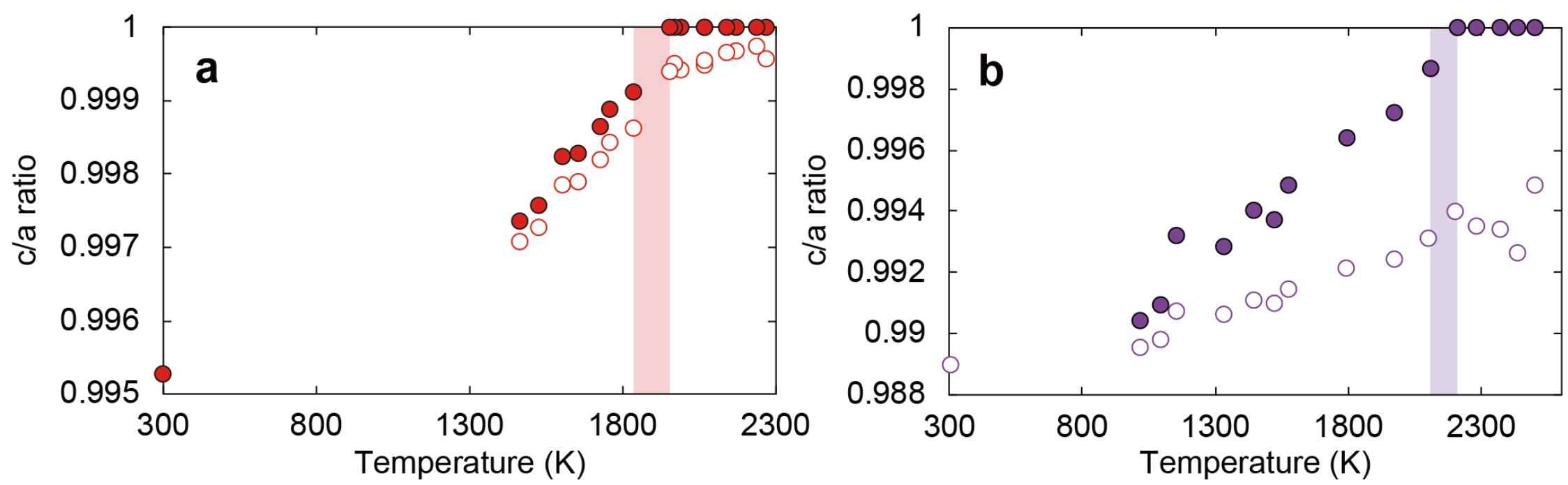

**Fig. S7. Temperature dependence of the *c*/*a* ratio of (a) $CaSiO_3$ davemaoite at ~60 GPa and (b) $CaSi_{0.75}Ti_{0.25}O_3$ davemaoite at ~ 50 GPa in run #1.** Open symbols show the raw data fitted using the *P4/mmm* tetragonal model, whereas filled symbols show the normalized c/a values, with the high-temperature plateau normalized to 1.

**Supplementary Table S1. Experimental conditions.**

| Composition | Run# | Pressure (GPa) | Temperature (K) |
|---|---|---|---|
| $CaSiO_3$ | 1 | 23–88 | 300–2360 |
| | 2 | 20–41 | 300–2010 |
| $Ca(Si_{0.75}Ti_{0.25})O_3$ | 1 | 20–68 | 300–2500 |
| | 2 | 14–72 | 300–2440 |

**Supplementary Table S2. Thermoelastic parameters for $CaSiO_3$ perovskite.**

| Thermoelastic parameters | This study (cubic) | This study (tetragonal) | This study (tetragonal, 300 K) | Ishii et al. (2026) | Sun et al. (2022) | Kawai & Tsuchiya (2014) | Noguchi et al. (2013) | Shim et al. (2000) | Wang et al. (1996) |
|---|---|---|---|---|---|---|---|---|---|
| $V_0$ (Å$^3$) | 45.54(16) | 45.49(7) | 45.49(6) | 45.43 | 45.60(40) | 46.21 | 45.8 | 45.58(3) | 45.58(3) |
| $K_{T0}$ (GPa) | 243(4) | 248(6) | 248(3) | 260.3 | 227(21) | 206.6 | 238 | 236(4) | 232(8) |
| $K'_{T0}$ | 4.05(16) | 4(fixed) | 4(fixed) | 3.70 | 4(fixed) | 4.41 | 4(fixed) | 3.90(20) | 4.8(3) |
| $\gamma_0$ | 1.47(3) | 1.22(20) | - | 1.772 | - | 1.567 | 2.8 | 1.92(5) | 1.7(fixed) |
| $q$ | 1.35(12) | 1.08(83) | - | 1.00 | - | 0.84 | 1.2 | 0.60(30) | 1.0(fixed) |
| $\theta_0$ (K) | 1100(fixed) | 1100(fixed) | - | 583.5 | - | 1100(fixed) | 1300 | 1000(fixed) | 1100(fixed) |

**Supplementary Table S3. Thermoelastic parameters for Ti25%-bearing $CaSiO_3$ perovskite.**

| Thermoelastic parameters | This study (cubic) | This study (tetragonal) | Chao et al. (2024) |
|---|---|---|---|
| $V_0$ (Å$^3$) | 48.9(2) | 45.60(40) | 29.68(5) |
| $K_{T0}$ (GPa) | 214(7) | 227(21) | 205(4) |
| $K'_{T0}$ | 4.35(69) | 4.00(30) | 4 (fixed) |
| $\gamma_0$ | 1.21(3) | - | - |
| $q$ | 1.20(72) | - | - |
| $\theta_0$ (K) | 900(fixed) | - | - |

**Supplementary References**

Adams, D.J., Oganov, A.R., 2006. *Ab initio* molecular dynamics study of $CaSiO_3$ perovskite at P−T conditions of Earth's lower mantle. Phys. Rev. B 73, 184106. https://doi.org/10.1103/PhysRevB.73.184106

Brown, J.M., Shankland, T.J., 1981. Thermodynamic parameters in the Earth as determined from seismic profiles. Geophys J Int 66, 579–596. https://doi.org/10.1111/j.1365-246X.1981.tb04891.x

Dziewonski, A.M., Anderson, D.L., 1981. Preliminary reference Earth model. Physics of the Earth and Planetary Interiors 25, 297–356. https://doi.org/10.1016/0031-9201(81)90046-7

Hirose, K., Takafuji, N., Sata, N., Ohishi, Y., 2005. Phase transition and density of subducted MORB crust in the lower mantle. Earth Planet. Sci. Lett. 237, 239–251. https://doi.org/10.1016/j.epsl.2005.06.035

Katsura, T., 2022. A Revised Adiabatic Temperature Profile for the Mantle. JGR Solid Earth 127, e2021JB023562. https://doi.org/10.1029/2021JB023562

Komabayashi, T., Hirose, K., Sata, N., Ohishi, Y., Dubrovinsky, L.S., 2007. Phase transition in $CaSiO_3$ perovskite. Earth Planet. Sci. Lett. 260, 564–569. https://doi.org/10.1016/j.epsl.2007.06.015

Kurashina, T., Hirose, K., Ono, S., Sata, N., Ohishi, Y., 2004. Phase transition in Al-bearing $CaSiO_3$ perovskite: implications for seismic discontinuities in the lower mantle. Physics of the Earth and Planetary Interiors 145, 67–74. https://doi.org/10.1016/j.pepi.2004.02.005

Lai, X., Zhu, F., Gao, J., Greenberg, E., Prakapenka, V.B., Meng, Y., Chen, B., 2022. Melting of the Fe-C-H System and Earth's Deep Carbon-Hydrogen Cycle. Geophys. Res. Lett. 49. https://doi.org/10.1029/2022GL098919

Ono, S., Ohishi, Y., Isshiki, M., Watanuki, T., 2005. In situ X-ray observations of phase assemblages in peridotite and basalt compositions at lower mantle conditions: Implications for density of subducted oceanic plate. J. Geophys. Res. 110, 2004JB003196. https://doi.org/10.1029/2004JB003196

Ono, S., Ohishi, Y., Mibe, K., 2004. Phase transition of Ca-perovskite and stability of Al-bearing Mg-perovskite in the lower mantle. Am. Mineral. 89, 1480–1485. https://doi.org/10.2138/am-2004-1016

Sagatova, D.N., Shatskiy, A.F., Sagatov, N.E., Litasov, K.D., 2021. Phase Relations in $CaSiO_3$ System up to 100 GPa and 2500 K. Geochem. Int. 59, 791–800. https://doi.org/10.1134/s0016702921080073

Shim, S.-H., Duffy, T.S., Shen, G., 2000. The stability and P-V-T equation of state of $CaSiO_3$ perovskite in the Earth's lower mantle. J. Geophys. Res. 105, 25955–25968. https://doi.org/10.1029/2000JB900183

Stixrude, L., Lithgow-Bertelloni, C., Kiefer, B., Fumagalli, P., 2007. Phase stability and shear softening in Ca Si O 3 perovskite at high pressure. Phys. Rev. B 75, 024108. https://doi.org/10.1103/PhysRevB.75.024108

Stixrude, L., Ronald Cohen, Yu, R., Krakauer, H., 1996. Prediction of phase transition in CaSi03 perovskite and implications for lower mantle structure. Am. Mineral. 81, 1293–1296.
Sun, N., Bian, H., Zhang, Y., Lin, J.-F., Prakapenka, V.B., Mao, Z., 2022. High-pressure experimental study of tetragonal CaSiO3-perovskite to 200 GPa. American Mineralogist 107, 110–115. https://doi.org/10.2138/am-2021-7913
Syracuse, E.M., Van Keken, P.E., Abers, G.A., 2010. The global range of subduction zone thermal models. Phys. Earth Planet. Inter. 183, 73–90. https://doi.org/10.1016/j.pepi.2010.02.004
Tateno, S., Hirose, K., Sata, N., Ohishi, Y., 2009. Determination of post-perovskite phase transition boundary up to 4400 K and implications for thermal structure in D″ layer. Earth Planet. Sci. Lett. 277, 130–136. https://doi.org/10.1016/j.epsl.2008.10.004
Thomson, A.R., Crichton, W.A., Brodholt, J.P., Wood, I.G., Siersch, N.C., Muir, J.M.R., Dobson, D.P., Hunt, S.A., 2019. Seismic velocities of CaSiO3 perovskite can explain LLSVPs in Earth’s lower mantle. Nature 572, 643–647. https://doi.org/10.1038/s41586-019-1483-x
Weck, G., Recoules, V., Queyroux, J.-A., Datchi, F., Bouchet, J., Ninet, S., Garbarino, G., Mezouar, M., Loubeyre, P., 2020. Determination of the melting curve of gold up to 110 GPa. Phys. Rev. B 101, 014106. https://doi.org/10.1103/PhysRevB.101.014106
Wu, F., Sun, Y., Wan, T., Wu, S., Wentzcovitch, R.M., 2024. Deep-Learning-Based Prediction of the Tetragonal → Cubic Transition in Davemaoite. Geophysical Research Letters 51, e2023GL108012. https://doi.org/10.1029/2023GL108012
Zhang, C., Yang, J.-Y., Sun, T., Zhang, H., Brodholt, J.P., 2025. Strong precursor softening in cubic $CaSiO_3$ perovskite. Proc. Natl. Acad. Sci. U.S.A. 122, e2410910122. https://doi.org/10.1073/pnas.2410910122